\pdfoutput=1

\documentclass[12pt,twoside,letterpaper]{article}
\usepackage{multicol}
\usepackage{paralist}
\usepackage{natbib}  %  NB: Can use bibpunct
\usepackage{ifpdf}

\usepackage[small,compact]{titlesec} %  Works, but only if the modified section
\ifpdf
\usepackage[pdftex]{graphicx}      %%% graphics for pdfLaTeX
\usepackage{epstopdf}
\usepackage[pdftex,dvipsnames,usenames]{color}
\usepackage[pdftex]{thumbpdf}      %%% thumbnails for pdflatex
\usepackage[pdftex,                %%% hyper-references for pdflatex
bookmarks=true,%                   %%% generate bookmarks ...
bookmarksnumbered=true,%           %%% ... with numbers
hypertexnames=false,%              %%% needed for correct links to figures !!!
breaklinks=true,%                  %%% break links if exceeding a single line
linkbordercolor={0 0 1}]{hyperref} %%% blue frames around links
\hypersetup{
pdfauthor   = {David Strickland <dks@pha.jhu.edu>},
pdftitle    = {Starburst Galaxies: Outflows of Metals and Energy into the IGM},
pdfsubject  = {Astro2010 White Paper},
pdfkeywords = {galaxies, starbursts, galactic winds, superwinds,
outflows, X-rays, point sources, luminosity functions, Chandra, Hubble,
Spitzer, IXO, galaxy formation, galaxy evolution},
}
\else
\usepackage[dvips]{graphicx}       %%% graphics for dvips
\usepackage[dvips,dvipsnames]{color}
\usepackage[ps2pdf]{thumbpdf}      %%% thumbnails for ps2pdf
\usepackage[ps2pdf,                %%% hyper-references for ps2pdf
bookmarks=true,%                   %%% generate bookmarks ...
bookmarksnumbered=true,%           %%% ... with numbers
hypertexnames=false,%              %%% needed for correct links to figures !!!
breaklinks=true,%                  %%% breaks lines, but links are very small
linkbordercolor={0 0 1}]{hyperref}%  %%% border-width of frames
\hypersetup{
pdfauthor   = {David Strickland <dks@pha.jhu.edu>},
pdftitle    = {Starburst Galaxies: Outflows of Metals and Energy into the IGM},
pdfsubject  = {Astro2010 White Paper},
pdfkeywords = {galaxies, starbursts, galactic winds, superwinds,
outflows, X-rays, point sources, luminosity functions, Chandra, Hubble,
Spitzer, IXO, galaxy formation, galaxy evolution},
pdfcreator  = {LaTeX with hyperref package},
pdfproducer = {dvips + ps2pdf}
}
\fi
\usepackage[rightcaption]{sidecap}

\bibpunct{(}{)}{;}{a}{\hspace{-0.5mm}}{,}

\protect
\protect        % LaTeX default is 0.70
\protect        % ..... ....... .. 0.20
\protect  %  Default 0.50

\newcommand{\captionfonts}{\small}
\makeatletter  %  Allow the use of @ in command names
\long\def\@makecaption#1#2{%
  \vskip\abovecaptionskip
  \sbox\@tempboxa{{\captionfonts #1: #2}}%
  \ifdim \wd\@tempboxa >\hsize
    {\captionfonts #1: #2\par}
  \else
    \hbox to\hsize{\hfil\box\@tempboxa\hfil}%
  \fi
  \vskip\belowcaptionskip}
\makeatother   %  Cancel the effect of \makeatletter

\newcommand\aj{{AJ}}%
\newcommand\araa{{ARA\&A}}%
\newcommand\apj{{ApJ}}%
\newcommand\apjl{{ApJ}}%
\newcommand\apjs{{ApJS}}%
\newcommand\aap{{A\&A}}%
\newcommand\mnras{{MNRAS}}%
\newcommand\pasj{{PASJ}}%
\newcommand\nat{{Nature}}%
\def\eg{{\rm e.g.\ }}

\def\ie{{\rm i.e.\ }}

\def\hi{${\rm HI}$}

\def\nii{[N{\sc ii}] \,}

\def\halpha{${\rm H}\alpha$}

\def\h50{\hbox{$h_{50}$\,}}
\def\spose#1{\hbox to 0pt{#1\hss}}
\def\ltsimm{\mathrel{\spose{\lower 3pt\hbox{$\sim$}}
        \raise 2.0pt\hbox{$<$}}}
\def\ltsim{$\mathrel{\spose{\lower 3pt\hbox{$\sim$}}
        \raise 2.0pt\hbox{$<$}}$}
\def\gtsimm{\mathrel{\spose{\lower 3pt\hbox{$\sim$}}
        \raise 2.0pt\hbox{$>$}}}
\def\gtsim{$\mathrel{\spose{\lower 3pt\hbox{$\sim$}}
        \raise 2.0pt\hbox{$>$}}$}

\def\fract#1/#2{\leavevmode\kern.1em                   % e.g. \fract 10/3
   \raise.5ex\hbox{\the\scriptfont0 #1}\kern-.1em
   /\kern-.15em\lower.25ex\hbox{\the\scriptfont0 #2}}
\def\deg{\hbox{$^\circ$}}
\def\arcm{\hbox{$^\prime$}}
\def\arcs{\arcm\hskip -0.1em\arcm}

\def\cm{{\rm\thinspace cm}}
\def\erg{{\rm\thinspace erg}}

\def\keV{{\rm\thinspace keV}}
\def\km{{\rm\thinspace km}}

\def\Lsol{\hbox{$\thinspace L_{\odot}$}}

\def\Msol{\hbox{$\thinspace M_{\odot}$}}
\def\pc{{\rm\thinspace pc}}
\def\s{{\rm\thinspace s}}

\def\yr{{\rm\thinspace yr}}
\def\pyr{\hbox{$\yr^{-1}\,$}}

\def\ergpcm3ps{\hbox{$\erg\cm^{-3}\s^{-1}\,$}}

\def\kmps{\hbox{$\km\s^{-1}\,$}}

\def\Lsolppc3{\hbox{$\Lsol\pc^{-3}\,$}}
\def\Msolppc3{\hbox{$\Msol\pc^{-3}\,$}}

\let\la=\ltsimm            % For Springer A&A compliance...
\let\ga=\gtsimm

\def\lt{\hbox{$\ltsimm$}}

\def\gt{\hbox{$\gtsimm$}}

\begin{document}

\begin{titlepage}
\renewcommand*{\pagestyle}{empty}
\begin{center}

\LARGE{\bf Starburst Galaxies: Outflows of Metals and Energy into the IGM}

\vspace*{0.5cm}

\Large{A White Paper for the Astro2010 Decadal Survey}

\vspace*{0.5cm}

\normalsize{
{\bf David K.~Strickland (Johns Hopkins University)}\\
Ann Hornschemeier (NASA/GSFC) \\
Andrew Ptak (Johns Hopkins University) \\
Eric Schlegel (University of Texas -- San Antonio) \\
Christy Tremonti (MPIA \& U.~Wisconsin) \\
Takeshi Tsuru (Kyoto University) \\
Ralph T\"{u}llmann (Harvard) \\
Andreas Zezas (Harvard) \\
} % end normalsize

\end{center}

\end{titlepage}

\newpage
\setcounter{page}{1}
\pagestyle{plain}
\begin{center}
%{\large{\bf IXO Stellar and Supernova Feedback Science }}
{\large{\bf Starburst Galaxies: Outflows of Metals and Energy into the IGM}}
%{\bf 2009 Decadal Survey White Paper}
\end{center}

\noindent{{\bf Key Question:} What is the contribution of mass, metals and 
energy from starburst galaxies to the Intergalactic Medium?}

\noindent{{\bf Summary of Present Knowledge:} 
%Starburst galaxies drive galactic-scale outflows or ``superwinds,''
%largely or solely powered by mechanical energy feedback from
%core-collapse supernovae and massive star stellar winds.
%Superwinds are a leading candidate mechanism for the 
%enrichment of the IGM and the loss of metals from low
%mass galaxies reflected in the galaxy mass-metallicity relationship.
%However we currently do not know how efficiently any starburst
%can eject metals via a superwind, nor the fraction of IGM metals
%carried there by superwinds, nor the type of galaxy dominating this
%process.
%We know that superwinds are
%common in both the low redshift and high redshift Universe. We have
%reasonably good measurements of the intensity of star formation 
%activity required to drive a superwind. 
%Furthermore moderately complex theoretical models of 
%(or incorporating) superwinds exist.
%Existing and near-future multi-wavelength 
%observational techniques are 
%progressing in providing a census of the multi-phase
%nature of superwinds from galaxies covering a broad range of galaxy mass, 
%although existing observational tools can not measure the kinematics 
%of the hottest gaseous phases of superwinds. 
%The hottest components of superwinds contain the majority of the energy and
%newly-synthesized metals in the superwind, and are thus crucial for
%assessing the global significance of 
%energy and chemical feedback from starbursts.
%Consequently \emph{we currently lack direct measurement of the rates at which
%starburst galaxies, as a function of galaxy mass, 
%eject gas, metals and energy into the IGM}. 
%}
Starburst galaxies drive galactic-scale outflows or ``superwinds''
that may be responsible for removing metals from galaxies and
polluting the Intergalactic Medium (IGM). Superwinds are powered
by massive star winds and by core collapse supernovae which collectively create
$T \la 10^{8}$~K bubbles of metal-enriched plasma within star forming
regions.  These over-pressured bubbles expand, sweep up cooler ambient
gas, and eventually blow out of the disk into the halo.  In the last
decade tremendous progress was made in mapping \emph{cool} entrained
gas in outflows through UV/optical imaging and absorption line
spectroscopy.  These studies demonstrated that superwinds are
ubiquitous in galaxies forming stars at high surface densities and that
the most powerful starbursts can drive outflows near escape velocity.
Theoretical models of galaxy evolution have begun to incorporate
superwinds, using various ad-hoc prescriptions based on our knowledge
of the cool gas.  However, these efforts are fundamentally impeded by
our lack of information about the \emph{hot} phase of these outflows.
\emph{
The hot  X-ray emitting phase of a superwind
contains the majority of its energy and
newly-synthesized metals, and given its high specific energy
and inefficient cooling it is also the
component most likely escape from the galaxy's gravitational potential well.
Knowledge of the chemical composition and velocity of the hot gas are
crucial to assess the energy and chemical feedback from a starburst.
}
A high priority for the next decade is to enable direct measurements
to be made
of the rates at which starburst galaxies of all masses eject gas,
metals, and energy into the IGM.

\noindent{{\bf Experimental Requirement Necessary to Answer Key Question:}
A high sensitivity X-ray imaging spectrometer capable of measuring
velocities in faint diffuse X-ray emission from plasmas in the temperature
range $10^{6} \la T (K) \la 10^{8}$, 
with a velocity accuracy of $\la 100$ km/s.
Such spectral resolution automatically allows detailed line-based
plasma diagnostics, and thus composition, energetics and flow rates
can be derived.
}

\section{Feedback between Stars, Galaxies and the IGM}
\label{sec:feedback}

We now know that galaxies and the IGM are intimately connected by 
flows of matter and energy. Both accretion onto galaxies and outflows
from galaxies or their central black holes link galaxies to the IGM in
what has been termed ``Cosmic Feedback.'' To obtain a deeper
physical understanding of either galaxy formation and 
evolution or the IGM requires 
that we better understand the physical processes that link them.

In many respects Cosmic Feedback is analogous to Stellar Feedback within
galaxies. Stars return both energy and matter back into the 
interstellar medium (ISM) from which they formed: 
either mechanically via metal-enriched stellar
winds (primarily from massive stars) and supernovae (SNe), or via ionizing
photons from hot massive stars. This combination of energetic and chemical
``feedback'' also plays an important, but
currently poorly understood, role in galaxy formation and 
evolution \citep[see \eg][]{kauffmann99,cen05,c_scannapieco08}. 

Mechanical feedback (stellar winds and SNe) is the primary physical
mechanism creating the hot phases of the ISM in star-forming galaxies 
(spiral, irregular and merging galaxies). The plasmas making up the
hot phases of the ISM have temperatures in the range 
$T = 10^{6}$ -- $10^{8}$ K and predominantly emit and absorb photons 
in the X-ray energy band from $E \sim 0.1$ -- 10 keV. 
Line emission (from highly-ionized ions of the astrophysically
%important elements oxygen, neon, magnesium, silicon, sulphur, and iron) 
important elements O, Ne, Mg, Si, S, and Fe) 
dominates the total emissivity of plasmas with temperatures 
$T \lt 10^{7}$ K, and at higher temperatures Ar, Ca 
and in particular Fe also produce strong lines.
Although the hot phases probably do not dominate the total mass
of the ISM in normal spiral galaxies (observationally the properties
of the hot phases are uncertain and remain a subject of vigorous ongoing
research) they dominate the energetics of the ISM and strongly
influence its
phase structure \citep{efstathiou00}.
X-ray observations are  a natural
and powerful probe of the composition and thermodynamic state of 
hot phases of the ISM in and around galaxies, and thus are also a powerful
tool for exploring the physics
of  feedback.

%- possible edit target, see christys suggestion
This White Paper focuses on the intersection of Cosmic and Stellar
Feedback. The intense star formation occurring in starburst galaxies
leads to the creation of energetic metal-enriched outflows
or superwinds that may pollute the IGM with metals and energy.
A high priority for the next decade is to enable direct measurements
to be made
of the rates at which starburst galaxies of all masses eject gas,
metals, and energy into the IGM.

\begin{figure}[t]
\centerline{%
\includegraphics[width=0.90\columnwidth]{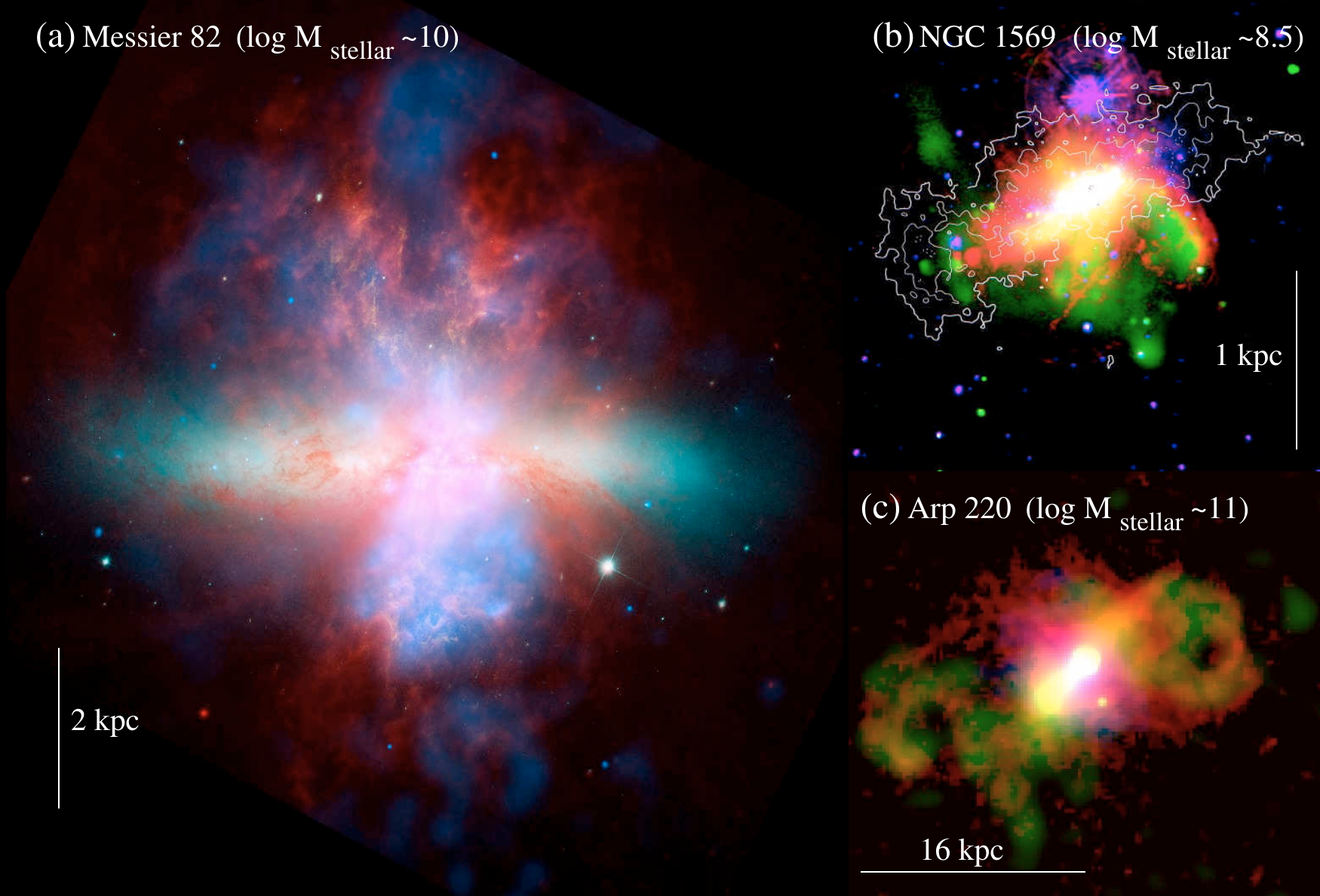}%
}
  \caption{Superwinds in nearby starburst galaxies of different mass.
   (a) Messier 82: The archetype of a starburst-driven superwind, as
   seen by the three NASA Great Observatories. Diffuse thermal X-ray
   emission as seen by {\it Chandra} is shown in blue. Hydrocarbon 
   emission at $8\mu$m from {\it Spitzer} is shown in red. Optical
   starlight (cyan) and \halpha+\nii~emission (yellow) are from {\it HST ACS}
   observations. 
   (b) The dwarf starburst NGC 1569.
   X-ray emission is shown in green, \halpha~emission in red, optical
   starlight in blue, and \hi~column density as white contours 
   \citep{martin02}.
   (c) The Ultra-luminous IR galaxy and merger-driven starburst Arp 220.
   X-ray emission is shown in green, \halpha~emission in red, and J-band 
   starlight in blue.
   }
  \label{fig:superwind_images}
\end{figure}

\section{Starburst-Driven Superwinds}

%The most important contribution IXO can make to our
%understanding of stellar and supernova feedback at galactic scales is
%to measure the velocity of the hot phases in starburst-driven
%galactic superwinds.

Superwinds are large-angle (opening angles $\gtsimm 30\deg$),
multi-phase, galactic-scale ($R \gtsimm 5$ -- 20 kpc) outflows
that have been observed in starbursting galaxies of all masses and
environments. Indeed, starburst galaxies with superwinds account 
for $\sim 20$\% of the high mass star formation in the local Universe,
having been observed in all galaxies 
where the average star formation rate per unit area
exceeds $\Sigma_{\rm SF} \ga  10^{-1} \Msol \pyr$ kpc$^{-2}$
\citep[see Fig.~\ref{fig:superwind_images}]{ham90,lehnert96b,veilleux05}. 
The UV-selected
star-forming galaxies at $z \sim 2$ -- 4 that may dominate the total
star formation density at these redshifts are known to drive powerful
winds that appear to be physically identical to local superwinds
\citep[see \eg][]{pettini01,shapley03}. 
Thus superwinds are a fundamental aspect of the Universe as we know it.

Superwinds are the strongest candidate for the cause of a number of
current astrophysical puzzles: the  origin of the galaxy mass-metallicity
relationship ($M-Z$), and more specifically the galaxy mass versus 
effective yield
relationship, $M-y_{\rm eff}$, which suggest that 
lower mass galaxies have
lost significant fractions of all the heavy elements ever 
created by their stars \citep{tremonti04}; the source of the
metal enrichment of the
IGM \citep[\eg][]{songaila97,simcoe06}; and the 
creation of $\sim 100$-kpc-scale holes in the IGM at
redshift $z \sim 3$ \citep{adelberger03}.

Although we have good reasons to suspect that superwinds
are solely or partly responsible for these phenomena we have
yet to robustly quantify their role. 
 We do not know the rates at which even local starbursts
eject gas, metals and energy, in particular because these are %probably
dominated by the hottest and most tenuous gaseous phases. Even if
winds are significant contributors to IGM enrichment we do not know what type
of galaxy dominates ejection (Which masses? What level of star formation?
Is galaxy environment important?),
and over what range of epochs this process is significant. There
is still much we don't know about the IGM baryon and metal budget.
The galaxy $M-y_{\rm eff}$ relationship is not a 
measure of the instantaneous metal
ejection efficiency for starbursts of a given mass, and
is potentially misleading because $y_{\rm eff}$ is only a sensitive
barometer of metal loss in gas rich systems with little star
formation subsequent to the burst \citep{dalcanton07}.
For example, some authors argue that only winds from the lowest mass galaxies
can reach the IGM \citep{ferrara00,keeney06}, while others 
argue that winds can indeed
escape from powerful starbursts in more massive 
galaxies \citep{strickland04b}.

Solving these problems requires the capability to directly measure the
pollution rate of the IGM with gas, metals, energy and momentum
by superwinds in galaxies covering a broad range of galaxy mass
and star formation activity. %At the very least we require the
%ability to measure the pollution rate in local ($z \ll 1$) starbursts
%covering a broad range in mass, and then
%extrapolate to superwinds at higher redshift.

\subsection{What We Currently Know About Superwinds}

Superwinds are driven by merged core-collapse SN ejecta and stellar winds,
which initially create a $T \sim 10^{8}$ K metal-enriched plasma within
the starburst region. This over-pressured gas expands and breaks out
of the disk of the host galaxy, converting
thermal energy into kinetic energy in a bi-polar outflow, which can potentially
reach a velocity of $3000 \kmps$.
This tenuous wind-fluid sweeps
up and accelerates cooler denser ambient
disk and halo gas. Radiation pressure may also play a role in accelerating
the dense entrained gas.

\begin{figure}[t]
\centerline{%
\includegraphics[width=0.85\columnwidth]{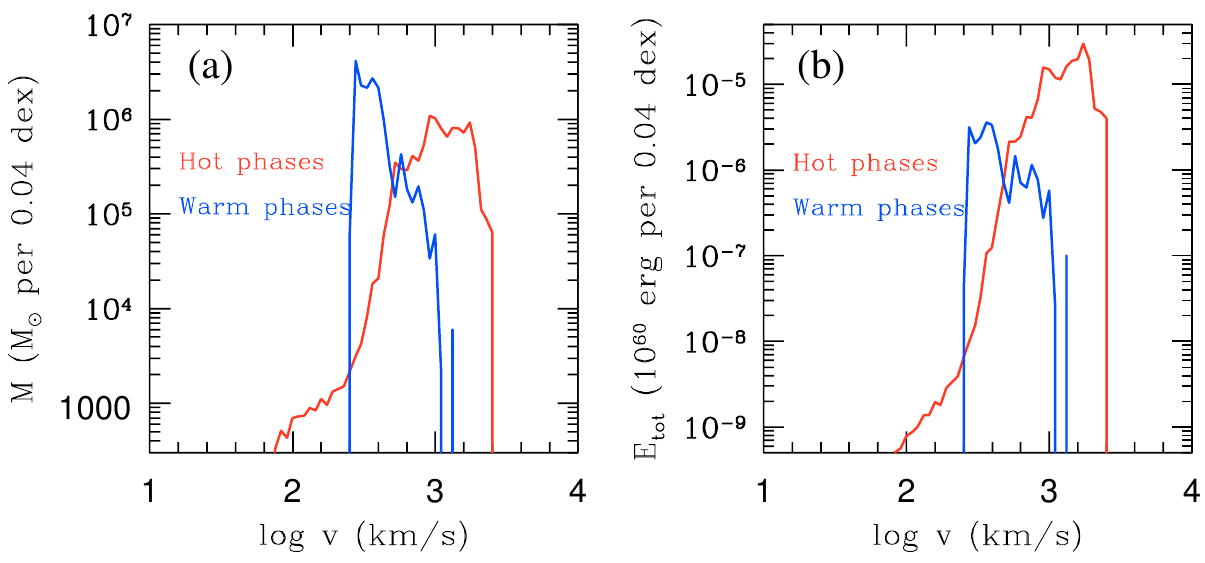}%
}
  \caption{Gas mass and total energy (thermal plus kinetic) as a function
  of velocity in a hydrodynamical simulation of a starburst-driven superwind.
  Current theoretical models for superwinds predict that the 
  hot phases ($6.3 \le \log T \le 8.3$, shown in red) have systemically
  higher outflow velocities than the warm neutral and ionized phases 
  ($3.8 \le \log T \le 4.2$, shown in blue) that are currently used to 
  measure velocities in superwinds. From \citet{strickland_dinge}.}
  \label{fig:cold_hot_velocities}
\end{figure}

Theoretical models predict that the entrained cool gas
is accelerated to lower velocities than the hot, metal-enriched, gas
\citep[\eg][see Figs.~\ref{fig:cold_hot_velocities} \& \ref{fig:cartoon}]{chevclegg,ss2000}. All existing 
observational velocity measurements of superwinds are
of the entrained cooler material, e.g. warm neutral and ionized gas
with outflow velocities in the range $200$ -- $1000 \kmps$ measured
using UV/optical emission and/or absorption lines 
\citep[see][]{shapley03,rupke05a,martin05}.
Although this material has velocities that are near galactic 
escape velocity, we expect that the hot phases are indeed escaping.

The theoretical models predict that the
majority of the energy ($90$\%) and metal content in superwinds
exists in the hot ($T \ge 10^{6}$ K) phases, with the kinetic energy
of such gas being several times the thermal energy.
Thus the metal-enriched phases are most likely to
escape into the IGM. This material has long been observationally
elusive, although we know believe we have firmly detected it
in X-ray emission \citep{strickland07,tsuru07}.

The current generation of X-ray 
telescopes offer spectral-imaging with a spectral resolution of order
$\Delta E \gt 100$ eV over the 0.3 -- 10 keV energy band 
(Chandra ACIS, XMM-Newton EPIC and Suzaku XIS). At this resolution the
emission lines are strongly blended, preventing the use of 
line-ratio-based spectral diagnostics. \emph{This spectral resolution is
too low to allow line shifts or broadening to be measured, preventing
any direct measurement of the velocity of the hot gasses.} Nor can 
existing X-ray gratings
be used to obtain higher spectral resolution, because the X-ray emission
is both spatially extended and faint.

Nevertheless spectral imaging
observations with {\it Chandra} and {\it XMM-Newton} detect thermal
X-ray emission from hot gas in superwinds extending out to
$5$ -- 30 kpc from the plane of edge-on starburst galaxies
\citep{strickland04a,tullmann06}.
Forward-fitting techniques do allow certain spectral parameters
such as temperature, emission integral ($n_{e}^{2} \, V$), and relative
elemental abundances (suggestive of enrichment by core-collapse SN)
to be crudely estimated
under the assumption of collisional ionization equilibrium.
Obtaining accurate plasma diagnostics, in particular of ionization state and
elemental abundances, will require higher spectral than existing
X-ray observatories can provide.

Although the average galaxy with a superwind at 
$z\sim3$ has a higher net star formation rate than the average local 
starburst, local starbursts with superwinds cover the same range of 
fundamental wind parameters such as warm gas
outflow velocity and mass flow rate, 
and star formation rate per unit area, as high redshift starbursts.
There are no significant obstacles that prevent us from applying the physics
of local superwinds to superwinds at the epoch of galaxy formation.

\section{Creating A Future for Superwind Studies}

Without direct measurements of the velocity of the hot gas in a superwind,
which can only be obtained with a high resolution X-ray imaging spectrometer, 
we can not know whether the hot metals created in the starburst
have sufficient energy to escape the galactic gravitational potential
well and reach the IGM. 

High spectral resolution would automatically allow the
use of line-ratio-based temperature and ionization state diagnostics,
and thus lead to more accurate elemental abundance determinations. 
Combined with velocity 
measurements we would obtain the mass, metal and energy flow rates that we
require to assess the impact and influence of superwinds.

High sensitivity will allow robust spectroscopy of the
very faint soft X-ray emission from the halos of starburst galaxies
and the accumulation of moderately-sized samples of local
galaxies covering meaningful ranges in parameter space.
We require measurements of gas
velocity and ejection rates in superwinds covering a range of 
different galaxy mass and 
star formation rate to assess which class of galaxies dominates the
metal enrichment of IGM in the present-day Universe, and in order
to relate the results to higher redshift galaxies and local
galaxy properties such as the galaxy $M-y_{\rm eff}$
relationship.

%\begin{figure}[t]
\begin{SCfigure}[][t]
%%\centerline{% centerline cannot work with SCfigure
\includegraphics[width=0.45\columnwidth]{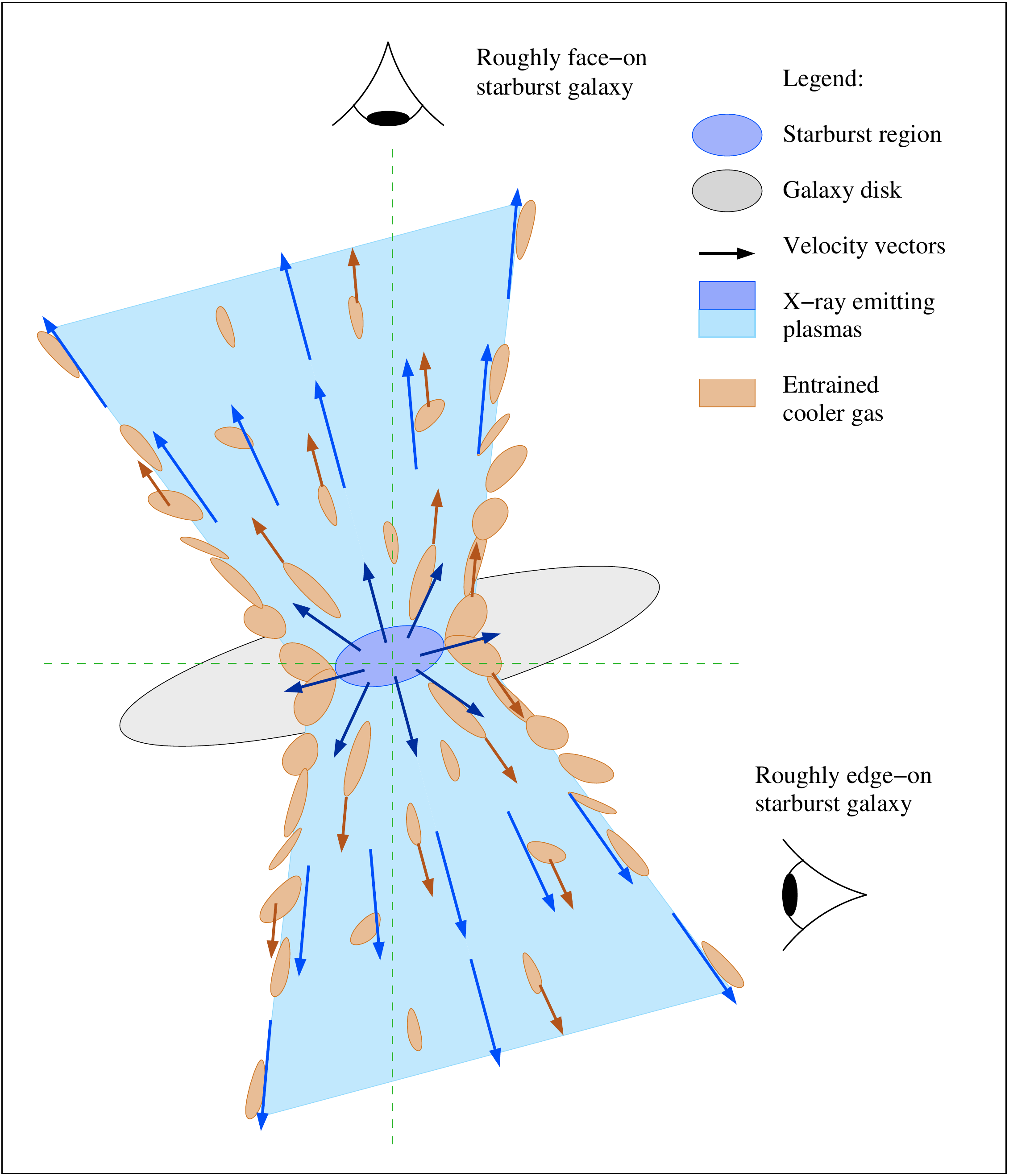}%
%%} %  end centerline
  \caption{Schematic diagram of a superwind viewed from different angles,
  with phase-dependent velocity vectors added to illustrate the line of
  sight velocity components expected. The combination of galaxy
  inclination and geometrical divergence within the wind leads to velocity
  shifts in the line centroids and line broadening or splitting. The 
  resulting velocity components
  along the line of sight are significant fractions of the 
  intrinsic velocity even
  in roughly edge-on starbursts.
  This diagram is simplified in that
  no acceleration or deceleration or change of geometry
  of the flow with position
  is shown. In practice we would look for such changes 
  (in particular in roughly edge-on systems) using spatially-resolved
  X-ray spectroscopy. [Figure best viewed in color.]
}
  \label{fig:cartoon}
\end{SCfigure}
%\end{figure}

This is not to suggest that advances in theoretical or other observational
capabilities are neither welcome nor necessary.
For example, UV/optical spectroscopy of the warm neutral and warm 
ionized phases (entrained gas) in superwinds provides the vital link that 
allows us to relate the properties of $z \ll 1$ superwinds
to the starburst-driven outflows at redshifts $z \ga 2$ 
\citep{shapley03}. Radio
and millimeter wavelength observations (\eg {\it ALMA}) 
constrain the molecular gas budget of winds and the nature of the starburst
regions where winds are launched. Nevertheless, without the X-ray
spectroscopic capability to
measure the ``true'' velocity of superwinds such progress would
be futile.

\section{Measuring Velocities in the X-ray-emitting Gas of a Superwind}
\label{sec:measuring}

%\subsection{Measuring Wind Velocities using Soft X-ray Emission}
%\label{sec:measuring:soft}

%\begin{figure}[t]
\begin{SCfigure}[][t]
%%\centerline{% centerline cannot work with SCfigure
\includegraphics[width=0.65\columnwidth]{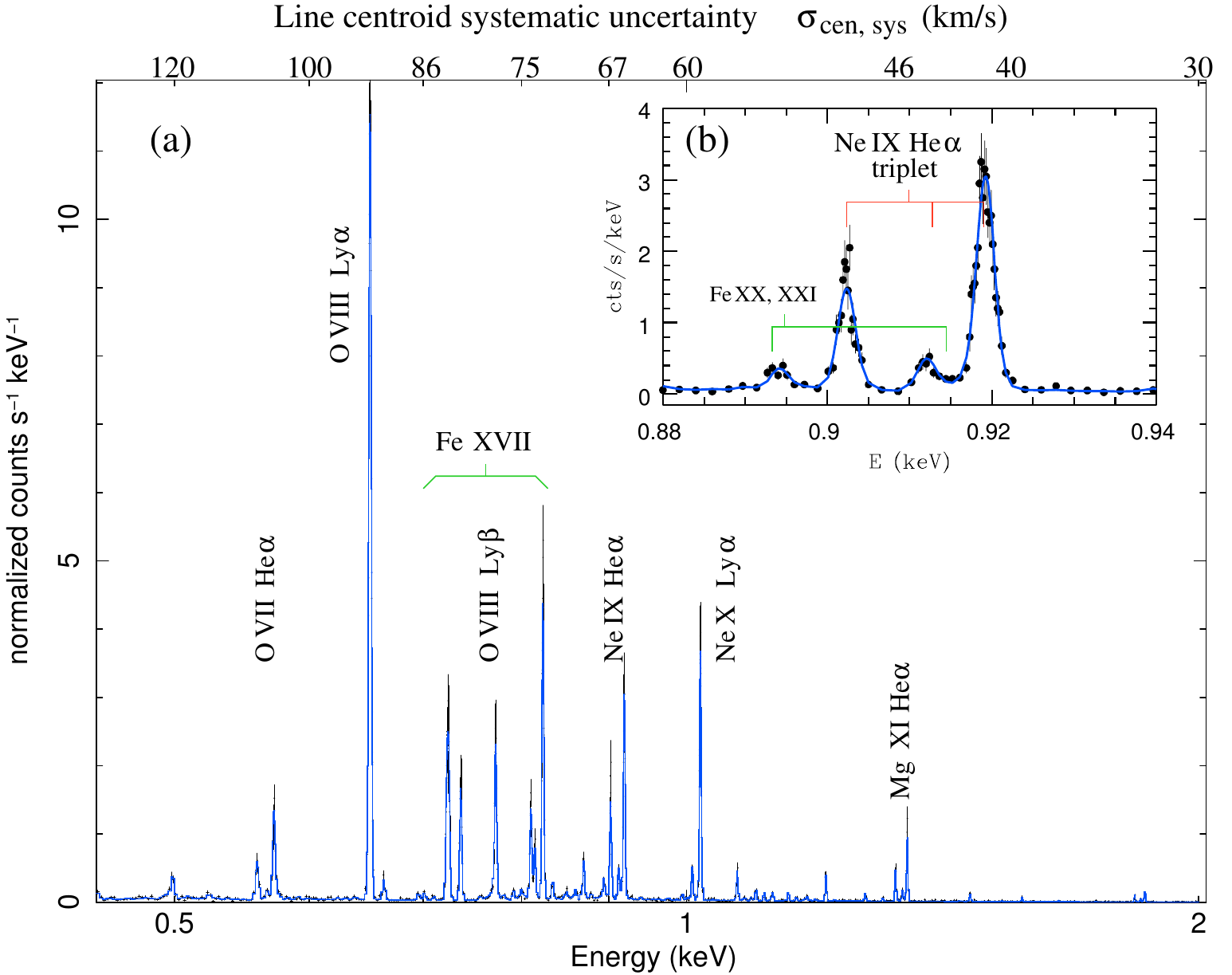}%
%%} %  end centerline
  \caption{(a) A simulated spectrum of a small 
region within a superwind based on the {\it IXO} XMS microcalorimeter
 instrument, illustrating the relative strength of the 
line emission. The simulation is of a supernova II-enriched
$kT=0.4$ keV thermal plasma containing 
$\sim 16000$ counts in the $E=0.3$ -- 2 keV energy band. The 
uncertainty in the line centroids due to the $\pm{0.2}$ eV 
systematic uncertainty in the absolute XMS energy scale is shown in terms
of the associated velocity uncertainty.
(b) A close-up of the
spectrum around the energy of the Ne IX He$\alpha$ triplet.%, illustrating how 
%well resolved the lines are
%and the small error bars on line and continuum emission.
}
  \label{fig:spectrum}
\end{SCfigure}
%\end{figure}

The remainder of this White Paper discusses the experimental accuracy
needed to achieve these goals, in particular velocity measurements, 
in comparison to the expected capabilities
of the {\it International X-ray Observatory} ({\it IXO}).
We consider only one of the possible observational
methods of measuring hot gas velocities in superwinds with {\it IXO},
specifically soft X-ray emission-line spectroscopy using the X-ray
Microcalorimeter System (XMS).
This method is most closely related in method to traditional 
observational studies
of the soft X-ray emitting plasmas in the halos of starbursts 
\citep[\eg][]{grimes05,tullmann06}.
The results presented here are derived and described in greater detail in the 
\href{http://proteus.pha.jhu.edu/~dks/Science/DecadalWP/index.html}{Technical Supplement}\footnote{The Technical Supplement can be accessed at 
\url{http://proteus.pha.jhu.edu/~dks/Science/DecadalWP/index.html}, or
by following the links in the PDF version of this White Paper.},
which also discusses alternative methods of measuring wind velocities in the
soft and hard X-ray bands.

What velocities do we expect for the X-ray emitting plasma in superwinds?
\begin{compactitem}
\item Gas motions must be comparable to galaxy escape velocities
to be significant in ejecting metals, where $v_{\rm esc} 
\sim 2$ -- $3 \times v_{\rm rot}$ (the rotational velocity of the host galaxy).
\item Theoretical models of superwinds 
predict high velocities (up to $3000 \kmps$) in gas with
$T \ga 10^{6}$ K, and that this material moves 
 significantly faster than the warm neutral and ionized medium in winds 
(\eg see Fig.~\ref{fig:cold_hot_velocities}). 
\item Even if this existing theory is completely wrong, for a wind to
exist the
hot gas velocity must be comparable or higher than the sound speed 
in the X-ray emitting gas, for which we have existing temperature 
measurements from {\it Chandra} and {\it XMM-Newton}. Thus
$v_{\rm HOT} \ga c_{\rm s} \sim 360 (kT_{\rm X}/0.5 \keV)^{0.5} \kmps$.
%even in dwarf starbursts.
\item Observed line-of-sight (LOS) velocities will be lower than the
intrinsic velocity, but even in the case of roughly edge-on
starbursts (e.g. M82) the geometry typically only decreases
the LOS velocity by a factor $\sim2$ from the intrinsic velocity
(see Fig.~\ref{fig:cartoon}).
\end{compactitem}

Outflows alter both the mean energy (\ie the line centroid) and width of the 
strong emission lines that dominate the soft thermal X-ray emission from
superwinds (Fig.~\ref{fig:spectrum}). We expect
LOS line shifts and line broadening in the range several hundred to a
thousand kilometers per second, hence requiring measurements 
accurate to $\la 100 \kmps$.

We find that the {\it IXO} XMS is capable of obtaining high 
quality X-ray spectra ($> 10^{4}$ counts) for even the faintest 
currently known sub-regions of superwinds in reasonable exposures 
($t_{\rm exp} \le 100$ ks). With such spectra we can measure 
\emph{individual} line centroid shifts with uncertainties $\sigma_{\rm cen} 
\sim 50 - 100$ km/s (68.3\% confidence), and line widths with uncertainties of 
$\sigma_{\rm FWHM} \sim 100$ km/s (this by fitting all lines in a 
single spectrum).
In fact it is calibration uncertainties that 
limit the line centroid measurement to a net accuracy of
 $\sigma_{\rm cen} \sim 50 - 100$ km/s.
 Note that both line widths and line centroids 
can be measured accurately to a fraction ($\sim10$\%) of the 
nominal instrument resolution of $\Delta E \approx 2.5$ eV (FWHM).

The IXO XMS provides its highest spectral resolution  over a 4 arcmin$^{2}$
field of view, with a spatial resolution $\sim 5\arcs$. This would allow
multiple high-quality spectra from different regions of a nearby superwind to
be accumulated simultaneously. This is advantageous as it allows
multiple sanity checks on the individual measurements, and 
the use of position-velocity 
diagrams to probe possible acceleration or deceleration in the wind
\citep[as done in the optical for superwinds, \eg][]{shopbell98}.

{\bf Observations using {\it IXO} of a sample of $\sim 30$
local starbursts, covering a suitably
broad range of galaxy mass ($8 \la \log M_{\rm stellar} (\Msol)
\la 11.5$, well matched to the local galaxy $M-y_{\rm eff}$ relationship
presented in \citealt{tremonti04}), are possible with a net exposure
of 1.3 Ms. Such a project will reveal whether starburst galaxies
are responsible for IGM enrichment and the galaxy mass-metallicity 
relationship.
}% end bf

%The X-ray calorimeter proposed for the {\it NeXT} mission 
%\citep[see \eg][]{mitsuda08} 
%will have sufficient sensitivity and spectral resolution to measure gas 
%velocities in the soft X-ray band of one or two of the very brightest 
%local starbursts --- the edge-on spiral
%galaxies M82 and (possibly) NGC 253. 
%Such observations would be historic as the
%first measurements of hot gas velocities in superwinds, but would not satisfy 
%our need to measure wind velocities in a statistically significant sample of 
%starburst galaxies covering a broad range in galaxy mass. Achieving this
%goal requires the combination of spectral resolution and sensitivity that
%only {\it IXO } can provide.

%----------------------------------------------------------------------
\begin{multicols}{2}
%\bibliographystyle{apj}
%%\bibliographystyle{shortref_nonum}
%%\bibliography{dks_refs}

\end{multicols}

\end{document}